\documentclass[aps,twocolumn,superscriptaddress,showpacs]{revtex4-1}
\usepackage[T1]{fontenc}
\usepackage{graphicx}
\usepackage{bm}
\usepackage{dcolumn}

\usepackage{epstopdf}
\usepackage{ulem}
\usepackage{siunitx}
\usepackage[ngerman]{babel}
\addto\captionsngerman{}  
\addto\captionsngerman{}  
\usepackage{natbib}
\usepackage{tabularx}
\newcolumntype{L}[1]{>{\raggedright\arraybackslash}p{#1}} 
\newcolumntype{C}[1]{>{\centering\arraybackslash}p{#1}} 
\newcolumntype{R}[1]{>{\raggedleft\arraybackslash}p{#1}} 

\usepackage{array}
\usepackage{textgreek}

\usepackage{multirow}

\begin{document}

\bibliographystyle{apsrev}

\title{Lattice dynamics and polarization-dependent phonon damping in $\boldsymbol{\alpha}$-phase FeSi$_{\textbf{2}}$  nanoislands}

\author{J. Kalt}
 \affiliation{Laboratory for Applications of Synchrotron Radiation, Karlsruhe Institute of Technology, \textit{D-76131} Karlsruhe, Germany}
 \affiliation{Institute for Photon Science and Synchrotron Radiation, Karlsruhe Institute of Technology, \textit{D-76344} Eggenstein-Leopoldshafen, Germany}

\author{M. Sternik}
 \affiliation{Institute of Nuclear Physics, Polish Academy of Sciences, \textit{PL-31342} Krak\'{o}w, Poland}

\author{B. Krause}
 \affiliation{Institute for Photon Science and Synchrotron Radiation, Karlsruhe Institute of Technology, \textit{D-76344} Eggenstein-Leopoldshafen, Germany}

\author{I. Sergueev}
 \affiliation{Deutsches Elektronen-Synchrotron, \textit{D-22607} Hamburg, Germany}

\author{M. Mikolasek}
 \affiliation{ESRF - The European Synchrotron, \textit{F-38000} Grenoble, France}

\author{D. Bessas}
 \affiliation{ESRF - The European Synchrotron, \textit{F-38000} Grenoble, France}

\author{O. Sikora}
 \affiliation{Institute of Nuclear Physics, Polish Academy of Sciences, \textit{PL-31342} Krak\'{o}w, Poland}

\author{T. Vitova}
 \affiliation{Institute for Nuclear Waste Disposal, Karlsruhe Institute of Technology, \textit{D-76344} Eggenstein-Leopoldshafen, Germany}

\author{J. G\"ottlicher}
 \affiliation{Institute for Photon Science and Synchrotron Radiation, Karlsruhe Institute of Technology, \textit{D-76344} Eggenstein-Leopoldshafen, Germany}

\author{R. Steininger}
 \affiliation{Institute for Photon Science and Synchrotron Radiation, Karlsruhe Institute of Technology, \textit{D-76344} Eggenstein-Leopoldshafen, Germany}

\author{P. T. Jochym}
 \affiliation{Institute of Nuclear Physics, Polish Academy of Sciences, \textit{PL-31342} Krak\'{o}w, Poland}

\author{A. Ptok}
 \affiliation{Institute of Nuclear Physics, Polish Academy of Sciences, \textit{PL-31342} Krak\'{o}w, Poland}

\author{O. Leupold}
 \affiliation{Deutsches Elektronen-Synchrotron, \textit{D-22607} Hamburg, Germany}

\author{H.-C. Wille}
 \affiliation{Deutsches Elektronen-Synchrotron, \textit{D-22607} Hamburg, Germany}
 
 \author{A. I. Chumakov}
 \affiliation{ESRF - The European Synchrotron, \textit{F-38000} Grenoble, France}

\author{P. Piekarz}
 \affiliation{Institute of Nuclear Physics, Polish Academy of Sciences, \textit{PL-31342} Krak\'{o}w, Poland}

\author{K. Parlinski}
 \affiliation{Institute of Nuclear Physics, Polish Academy of Sciences, \textit{PL-31342} Krak\'{o}w, Poland}

\author{T. Baumbach}
 \affiliation{Laboratory for Applications of Synchrotron Radiation, Karlsruhe Institute of Technology, \textit{D-76131} Karlsruhe, Germany}
 \affiliation{Institute for Photon Science and Synchrotron Radiation, Karlsruhe Institute of Technology, \textit{D-76344} Eggenstein-Leopoldshafen, Germany}

\author{S. Stankov}
 \email{svetoslav.stankov@kit.edu}
 \affiliation{Laboratory for Applications of Synchrotron Radiation, Karlsruhe Institute of Technology, \textit{D-76131} Karlsruhe, Germany}
 \affiliation{Institute for Photon Science and Synchrotron Radiation, Karlsruhe Institute of Technology, \textit{D-76344} Eggenstein-Leopoldshafen, Germany}

\date{\today}

\begin{abstract}
We determined the lattice dynamics of metastable, surface-stabilized $\alpha$-phase FeSi$_2$ nanoislands epitaxially grown on the Si(111) surface with average heights and widths ranging from 1.5 to 20\,nm and 18 to 72\,nm, respectively. The crystallographic orientation, surface morphology and local crystal structure of the nanoislands were investigated by reflection high-energy electron diffraction, atomic force microscopy and X-ray absorption spectroscopy. The Fe-partial phonon density of states (PDOS), obtained by nuclear inelastic scattering, exhibits a pronounced damping and broadening of the spectral features with decreasing average island height. First-principles calculations of the polarization-projected Si- and Fe-partial phonon dispersions and PDOS enable the disentanglement of the contribution of the \textit{xy}- and \textit{z}-polarized phonons to the experimental PDOS. Modeling of the experimental data with the theoretical results unveils an enhanced damping of the \textit{z}-polarized phonons for islands with average sizes below 10\,nm.
This phenomenon is attributed to the fact that the low-energy \textit{z}-polarized phonons couple to the low-energy surface/interface vibrational modes. The thermodynamic and elastic properties obtained from the experimental data show a pronounced size-dependent behavior.
\end{abstract}


\maketitle

Nanostructures of transition metal silicides have a wide range of applications and constitute fundamental building blocks of current micro- and nanoelectronics \cite{Murarka_silicides_microelectronics,chen_silicides_microelectronics,burkov_silicide_thermoelectrics}. Among these compounds, FeSi$_2$ is particularly interesting since it is the only representative that forms metallic and semiconducting phases \cite{liang_nw_phase}.
The thermodynamic phase diagram of bulk FeSi$_2$ shows a transition of the room-temperature stable semiconducting $\beta$-phase to the high-temperature metallic  $\alpha$-phase at 950$\,^\circ$C  \cite{starke_phases}.
A large number of studies investigated the formation of iron silicide thin films on silicon substrates and revealed two metastable metallic phases (\textit{s}- and $\gamma$-FeSi$_2$) with cubic structure and lattice parameters close to the value of silicon (e.g. \cite{kaenel_sphase_prb,alvarez_phase_Si111,kataoka_phase_Si111,nakano_phase_Si001}). 
Up to a critical thickness, the formation of the lattice-matched metastable phases is energetically favorable over the formation of strained $\beta$-FeSi$_2$. 
In a similar manner, the tetragonal high-temperature phase $\alpha$-FeSi$_2$  can be stabilized at room temperature in epitaxial nanostructures by deposition of a few Fe monolayers on the Si surface \cite{chevrier_alpha_rheed,berbezier_TEM, jedrecy_alpha_TEM, sauvage_Fe_Si, stocker_alpha_Si111, kataoka_phase, sirotti_alpha_beta}.
Several experimental and theoretical studies investigated the  magnetic \cite{miiller_Fe_valence,dascalu_supermagnetic,zhandun_SCM} and electronic \cite{sirotti_alpha_beta,kurganskii_electronic,tarasov_etrans,sandalov_e_correlation} properties of $\alpha$-FeSi$_2$ nanostructures. 
The discovery of superparamagnetic behavior in nanoislands and nanostripes \cite{tripathi_islands,tripathi_stripes}, the indication of a ferromagnetic-semiconductor-like behavior below 50\,K \cite{Cao_PRL} and the fabrication of $\alpha$-FeSi$_2$ nanobars \cite{xu_alpha_nanobars} and nanowires \cite{zou_alpha_wires} suggested applications of this material in nanoelectronics.

The collective vibrations of atoms in a crystal are characterized by the phonon dispersions and phonon density of states (PDOS) and play an important role for the properties of  materials. 
For instance, via the vibrational entropy they govern phase transitions, in semiconductors and insulators they are decisive for the heat transport, and through interactions with electrons and magnons they can affect other application-relevant properties. 
It is well known that reduction of the size of crystals to the nanometer length scale induces pronounced changes in the vibrational and thermodynamic properties due to broken translational symmetry at surfaces and interfaces  \cite{cuenya_Fe_clusters,cuenya_FePt_clusters,cuenya_Pt_nanoparticles,cuenya_Fe_particles,slezak_PRL,stankov_nanograins,stankov_nano_alloy,bozyigit_e_ph,Seiler_PRL,keune_if,Fe3Si_PRB,Sikora_fe3si,lazewski_Fe_surface}, epitaxial strain \cite{Stankov_PRL_Fe}, coupling to the surrounding \cite{shi_qd,cuenya_multilayer,sopu_ge_pdos}, or magnetic ordering \cite{spiridis_2d_3d}. 
Furthermore, at dimensions comparable to the phonon wavelengths quantization phenomena are predicted \cite{Sauceda_Au_nanoparticles,Piekarz_FePt_nanoalloy}.

Commonly observed effects in the PDOS are an enhancement of the number of phonon states in the low- and high-energy part and a broadening of the spectral features, compared to the bulk counterparts. 
This also applies for nanoislands, which exhibit several additional effects upon nanoscaling.
For example, in Fe nanoclusters a deviation from the Debye law was observed in the low-energy part of the PDOS, which is attributed to vibrations of low-coordinated surface and interface atoms \cite{cuenya_Fe_clusters}.
A high sensitivity of the vibrational properties to the core/shell structure was reported in nanoparticles consisting of an FePt core and a PtSi shell \cite{cuenya_FePt_clusters}.
In Pt nanoparticles a reduction of the bond length leads to stiffening of the crystal and an increase of the Debye temperature \cite{cuenya_Pt_nanoparticles}.
A study on EuSi$_2$ nanoislands revealed the existence of high-energy vibrational modes, which emerge at the interfaces of adjacent islands \cite{Seiler_PRL}. 
It was shown that the lattice softening commonly observed in nanostructures alters the interaction of phonons with electrons, magnons and other phonons \cite{pradip_EuO_prl,pradip_EuO_nscale,bozyigit_e_ph}. 
On the one hand this can lead to a deterioration of functionality of nanoelectronic devices, while on the other hand this can be beneficial for applications in superconductivity or thermoelectrics.
Moreover, the controlled modification of the lattice dynamics by nanoscaling can pave the way towards the design of nanostructures with tailored vibrational properties, which is the main objective of phonon engineering \cite{balandin_phononics_nano,balandin_phononics_lowd}.
Therefore, a comprehensive understanding of the lattice dynamics in nanostructures is of fundamental importance.

The experimental determination of the lattice dynamics of nanostructures, however, remains a challenge in modern solid state physics. 
Nuclear inelastic scattering (NIS) \cite{seto_prl_nis} has proven to be a unique method to overcome several problems connected to such measurements. 
It gives access to the element- and isotope-specific PDOS \cite{sturhahn_PRL_1995} and provides the high sensitivity required for the measurement of very small amounts of material present in nanostructures \cite{Stankov_PRL_Fe}.
The high penetration depth of the X-rays enables the measurement of thin buried layers, which is not feasible with other methods. 
Moreover, it was shown experimentally \cite{chumakov_anisotropic_nis_FeBO3} and described theoretically \cite{kohn_anisotropic_nis_theo} that NIS also enables the measurement of the polarization-dependent PDOS in anisotropic single-crystals. 
Employing NIS and \textit{ab initio} calculations the lattice dynamics of bulk $\beta$-FeSi$_2$ was studied comprehensively \cite{walterfang_beta_PDOS,liang_FeSi_abinitio,beta_PDR}. 
For $\alpha$-FeSi$_2$, however, only particular thermodynamic properties were predicted theoretically \cite{liang_FeSi_abinitio,acker_TDP_alpha}.

Here we present a combined experimental and theoretical  study of the lattice dynamics of surface-stabilized $\alpha$-phase FeSi$_2$ nanoislands.
Due to the specific orientation of the $\alpha$-FeSi$_2$ unit cell on the Si(111) surface, the NIS experiment provides access to lattice vibrations with projections along the main crystallographic directions. 
The Fe-partial PDOS exhibits a strong dependence on the size of the nanoislands.
A comparison of the experimental data with the \textit{ab initio} results reveals a stronger damping of the \textit{z}-polarized phonons  compared to \textit{xy}-polarized phonons for an average island height below 10\,nm.

The paper is organized as follows: in Sec. \ref{EXPERIMENT} the growth procedure and experimental methods are described, in Sec. \ref{THEORY} the \textit{ab initio} calculation and modeling details are given. Section \ref{Results} A presents the results of the structural investigation, in subsection B the results of the \textit{ab initio} calculations and the NIS experiment are discussed and in subsection C the results of the thermodynamic properties are given. The conclusions are included in Sec. \ref{Conclusions}.

\section{Experimental details}\label{EXPERIMENT}
Epitaxial $\alpha$-phase FeSi$_2$ nanoislands were grown on Si(111) substrates under ultra high vacuum (UHV) conditions (P$<$1$\times$10$^{-8}$\,Pa) in the UHV-Analysis lab at KIT. 
First, the substrate was degassed in UHV at 650$\,^\circ$C for 4\,h. Subsequently, the native SiO$_2$ layer was removed by heating the substrate two times to 1250$\,^\circ$C for 30 seconds. 
An atomic beam of high purity iron, enriched to 96\,\% in the M\"ossbauer-active isotope $^{57}$Fe, was supplied from an electron beam evaporator. 
The coverage was controlled by a quartz oscillator with an accuracy of 10\,\%.
The samples were grown by depositing a certain amount  of iron $\theta_{Fe}$ onto the Si(111) substrate heated to the growth temperature $T_G$, a process known as reactive deposition epitaxy (RDE), which is commonly used for the growth of iron silicide nanostructures  (e.g. \cite{chevrier_alpha_rheed,minami2002rheed,berbezier_TEM,matsumoto_SPE_RDE,ohira2008_wires,gonzalez2010_RDE}). 
Six samples, hereinafter referred to as S1-S6, were prepared, characterized and investigated. 
Details of the growth and experimental conditions are summarized in Table\,\ref{tab:samples}. 
Directly after the growth process S1 was annealed at $T_A$\,=\,770$\,^\circ$C for $t_A$\,=\,2\,h to examine possible effects of annealing on the crystal structure and the lattice dynamics. 
The temperature values are measured with an accuracy of $\pm$\,10$\,^\circ$C. All measurements described in the following were conducted at room temperature. The crystal structure was investigated with reflection high-energy electron diffraction (RHEED). Afterwards the samples were transferred under UHV condition to an {\sc Omicron Large Sample} scanning probe microscope operated in a non-contact atomic force microscopy mode to determine the surface topography.
S1, S3, S5 and S6 were subsequently capped with 4\,nm of amorphous Si sputtered at room temperature in a chamber \cite{sputter_krause} with a base pressure of P$<$1$\times$10$^{-6}$\,Pa also connected to the UHV-Cluster. 
The flux of the sputter gas Ar was 0.8\,sccm, corresponding to a pressure of 0.36 Pa.

\begin{table}[b]
\renewcommand{\arraystretch}{1.2}
 \begin{tabular}{|c|c|c|c|c|c|}
 \hline
Sample   & $\theta_{Fe}$ [\AA] &    $T_G$  [$\,^\circ$C]    &   $T_A$  [$\,^\circ$C] 	        &  $t_A$ [h] & NIS exp. 		    \\                   
\hline
S1		 &   2.2(2)			 &     700(10)   					& 770(10)					& 	2		 & Si cap 		  		 \\
S2 	     &   2.2(2)			 &     700(10)				   		&		-					&		- 	 & \textit{in situ}   \\
S3       &   0.6(1)			 &     700(10)			   			& 		-					&		- 	 & Si cap 		  		  \\
S4    	 &   2.2(2)			 &     500(10)			   			&		-					&		-	 & \textit{in situ}   \\
S5    	 &   0.6(1)			 &     650(10)			   			&		-					&		-	 & Si cap 		  		  \\
S6       &   0.6(1)			 &     500(10)			   			&		-	     			& 		 -	 & Si cap 		  		  \\
\hline
 \end{tabular}
 \caption{Overview of the investigated samples. $\theta_{Fe}$ stands for the deposited amount of $^{57}$Fe, $T_G$  for the growth temperature,  $T_A$ for the annealing temperature and $t_A$ for the annealing time. The last column denotes if the sample was capped with Si or measured \textit{in situ} during the NIS experiment.  
}  \label{tab:samples}
\end{table}

The Fe-partial PDOS was obtained \cite{evaluation_pdos} from nuclear inelastic scattering experiments
performed at the Dynamics Beamline P01 \cite{p01} at PETRA III and the Nuclear Resonance Beamline ID18 \cite{id18} at the ESRF. Samples S2 and S4 were transferred to the beamlines and measured \textit{in situ}, i.e. under UHV condition (P$<$5$\times$10$^{-7}$\,Pa) in a dedicated UHV chamber \cite{ibrahimkutty_chamber}. At both beamlines the measurements were performed at grazing-incidence geometry with an incidence angle\,<\,0.2$^{\circ}$ and an X-ray beam with dimensions of 1.5\,mm\,$\times$\,0.01 mm (h\,$\times$\,v, FWHM). The energy resolution for the photons with an energy of 14.4\,keV was 0.7 meV at ID18 (S1, S2) and 1.1 meV at P01 (S3-S6). After the \textit{in situ} experiment  samples S2 and S4 were transferred back under UHV conditions and also covered with a 4\,nm thick Si layer. 

Additionally, the local crystal structure of the FeSi$_2$ samples was characterized by Fe K-edge X-ray absorption spectroscopy at the SUL-X beamline of the synchrotron radiation source KARA at KIT.
After calibration with an $\alpha$-Fe metal foil to 7112\,eV  (Fe\,K-edge), fluorescence emission of the samples was recorded up to $k\,=\,14\,$\AA$^{-1}$. 
The EXAFS spectra were obtained with a beam-to-sample-to-detector geometry of 45$^{\circ}$/45$^{\circ}$ using a collimated X-ray beam of about 0.8\,mm\,$\times$\,0.8\,mm, or focused X-ray beam with  0.35\,mm\,$\times$\,0.15\,mm (h\,$\times$\,v, FWHM) at the sample position.

\section{Theoretical details}\label{THEORY}

In order to attain a comprehensive understanding of the lattice dynamics of $\alpha$-FeSi$_2$, first-principles calculations were performed within the density functional theory (DFT) implemented in the VASP code \cite{vasp1,vasp2}, employing the generalized gradient approximation \cite{PBE1,PBE2}. 
The interaction between ions and electrons was described using the projector augmented-wave method \cite{PAW1,PAW2}, with plane waves basis expanded up to a cutoff energy of 400\,eV. 
The configurations Si(\textit{s${\,^{2}}$p${\,^{2}}$}), and Fe(\textit{d${\,^{7}}$s${\,^{1}}$}) were treated as valence electrons.  
The $\alpha$-FeSi$_{2}$ phase was modeled by imposing the symmetry restrictions of the tetragonal $P4/mmm$ space group on the crystal structure. 
The primitive cell contains one formula unit, i.e. 3 atoms, with two nonequivalent positions: Fe placed in (0,0,0) and Si in (0.5,0.5,\textit{z}).  
Calculations were carried out in a $4\times4\times2$  supercell containing 64 Si and 32 Fe atoms using the $2\times2\times2$ Monkhorst–Pack grid of k-points. 
The convergence criteria for the total energy and internal forces of 10$^{-8}$\,eV and 10$^{-6}$\,eV \AA$^{-1}$, respectively,  were applied.
After the geometry relaxation we obtained the lattice parameters $a\,=\,2.702$\,\AA ~ and $c\,=\,5.140$\,\AA, and the internal atomic position $z\,=\,0.2725$.  
The calculated lattice constants are in very good agreement with the published experimental  ($a\,=\,2.68$\,\AA\, and $c\,=\,5.13$\,\AA) \cite{exper} and theoretical data ($a\,=\,2.70$\,\AA\, and $c\,=\,5.13$\,\AA) \cite{sandalov_e_correlation}. 
All calculations were performed assuming ferromagnetic order of the Fe atoms, however, the obtained magnetic moments are negligibly small ($\sim$\,0.01\,$\mu _B$). 
The phonon dispersion relations and PDOS were calculated at 0\,K using the direct method \cite{phonon1} incorporated into the PHONON program \cite{phonon2}. 
This method utilizes the DFT calculated Hellmann-Feynman forces generated by displacing the nonequivalent atoms from their equilibrium positions.

The EXAFS spectra were processed and modeled using the {\sc ATHENA} and {\sc ARTEMIS} programs included in the {\sc IFEFFIT} package \cite{ravel_athena}.
The spectra were weighted by $k$\,=\,1,\,2, and 3\,\AA$^{-1}$  within the k ranges given in Table \ref{tab:tab_EXAFS}. 
Hanning windows and $dk$\,=\,2 were used. 
A shell-by-shell approach was applied to model the data in real space within a range of 1.0\,-\,2.7\,\AA. 
Multiple scattering paths do not contribute in the modeled R region.
The crystal structure of $\alpha$-FeSi$_2$ was used to calculate the single scattering paths. 
The amplitude reduction factor was set to 0.7 and was fixed during the fitting process. 
It was obtained by modeling the EXAFS spectrum of an $\alpha$-Fe foil measured at the same experimental conditions. The Debye-Waller parameters for Si were free, whereas the values for Fe obtained from the NIS experiment were fixed  during the fit.

\section{Results and discussions}\label{Results}

\subsection{Structural investigation}

\begin{figure}[b]
\includegraphics[width=0.99\columnwidth]{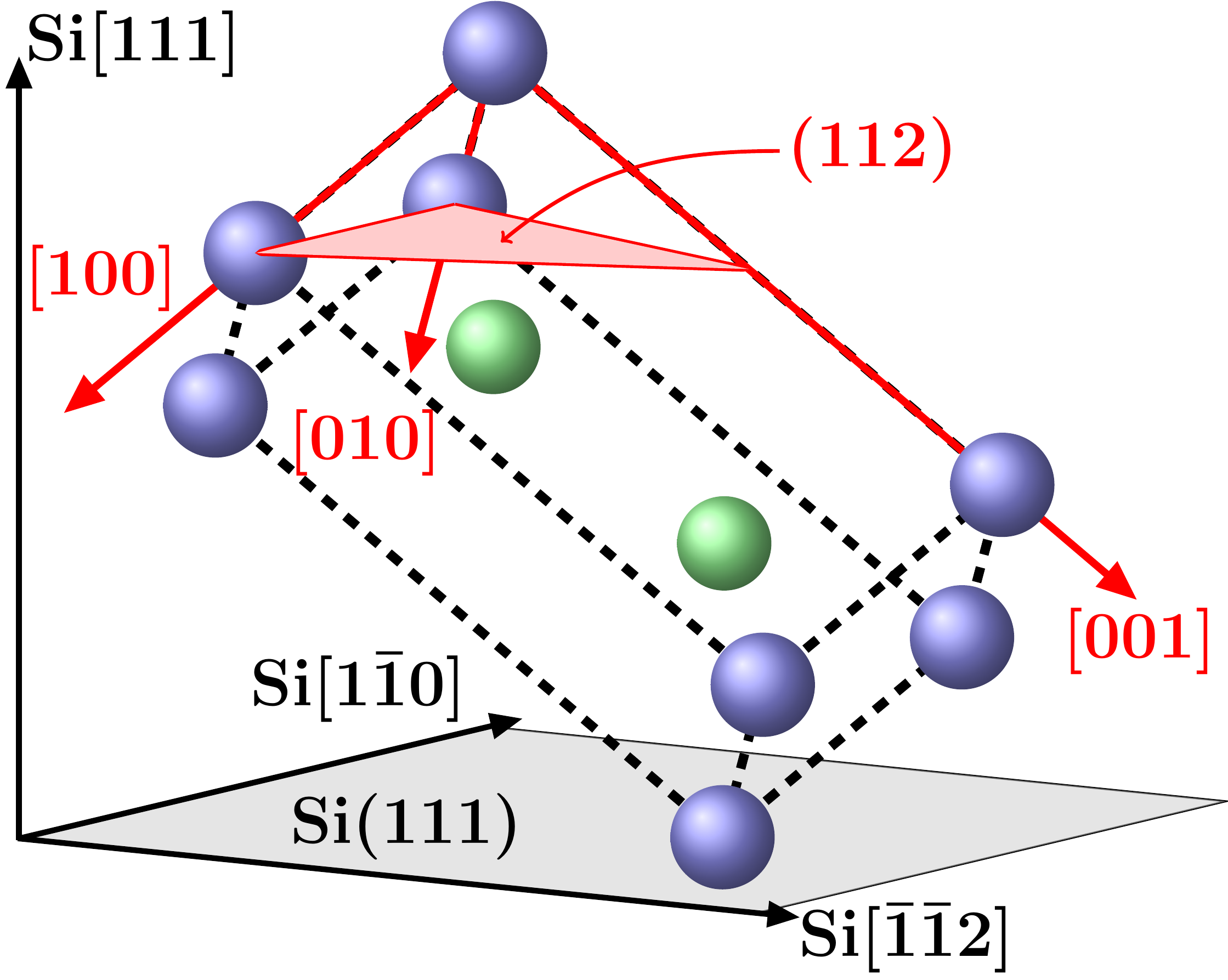}
\caption{Orientation (i) (see text) of the $\alpha$-FeSi$_2$ unit cell on the Si(111) surface. Directions and planes related to Si ($\alpha$-FeSi$_2$) are given in black (red). Fe atoms are depicted in blue, Si atoms in green. 
}
\label{fig:surface}
\end{figure}

\begin{figure}[h!]
\includegraphics[width=0.91\columnwidth]{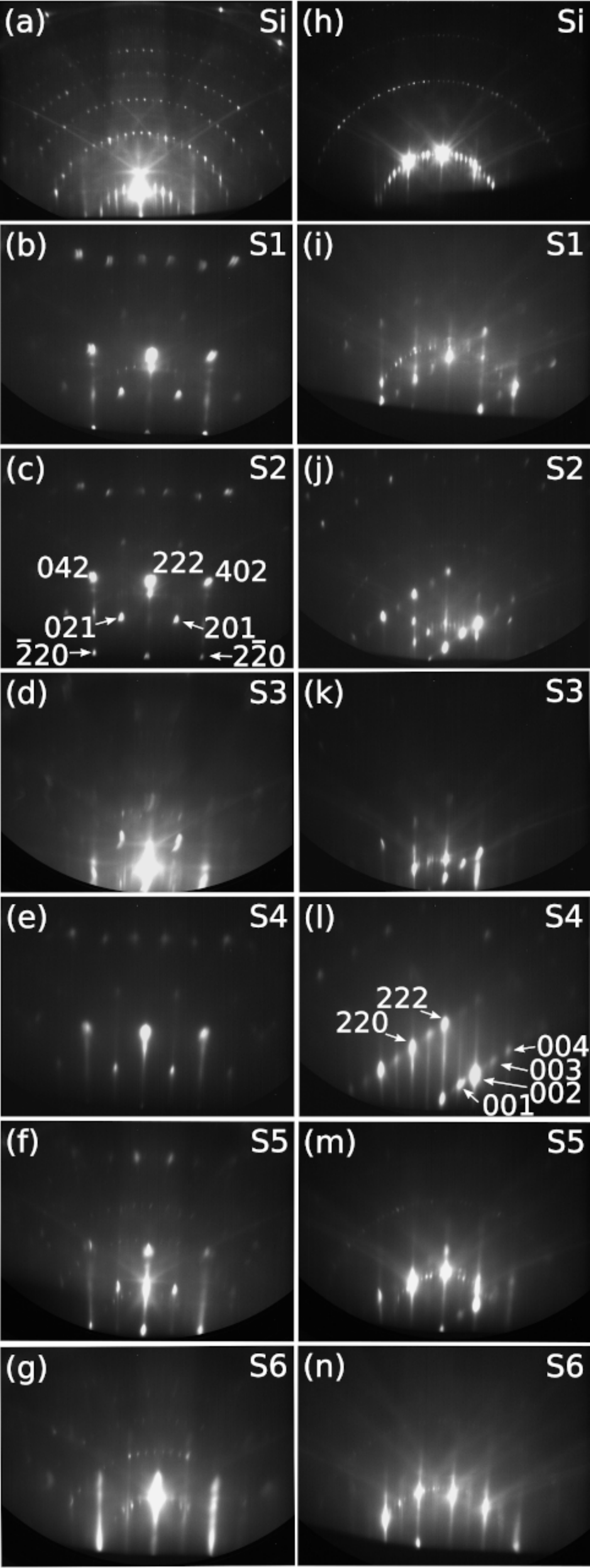}
\caption{RHEED patterns of the Si substrate (a,h) and the investigated samples (b-g, i-n) obtained with E\,=\,28\,keV along Si$\langle 11\bar{2}  \rangle$ (a-g) and Si$\langle\bar{1}$10$\rangle$ (h-n). In (c) and (l) the indexes of the reflections are given following \cite{jedrecy_alpha_TEM}.}
\label{fig:RHEED}
\end{figure}

The epitaxial growth of $\alpha$-FeSi$_2$  on the Si(111) surface has previously been investigated by grazing-incidence X-ray diffraction \cite{stocker_alpha_Si111},  electron microscopy \cite{berbezier_TEM,sauvage_Fe_Si,jedrecy_alpha_TEM} and combined RHEED and grazing-incidence X-ray diffraction \cite{chevrier_alpha_rheed}. 
It has been found that the tetragonal $\alpha$-FeSi$_2$ unit cell is oriented with its (112) plane parallel to the Si(111) plane ($\alpha$-FeSi$_2$(112)$\vert\vert$Si(111)). 
The $\alpha$-FeSi$_2$ unit cell can be accommodated on the Si(111) surface in three different domain orientations rotated by 120$^\circ$: (i) $\alpha$-FeSi$_2$[$\bar{1}$10]$\vert\vert$Si$[1\bar{1}$0] and $\alpha$-FeSi$_2$[$\bar{1}\bar{1}$1]$\vert\vert$Si$[\bar{1}\bar{1}$2], (ii) $\alpha$-FeSi$_2$[$\bar{2}$01]$\vert\vert$Si$[1\bar{1}$0] and $\alpha$-FeSi$_2$[2$\bar{4}$1]$\vert\vert$Si$[\bar{1}\bar{1}$2], as well as (iii) $\alpha$-FeSi$_2$[0$\bar{2}$1]$\vert\vert$Si$[1\bar{1}$0] and $\alpha$-FeSi$_2$[4$\bar{2}$1]$\vert\vert$Si$[\bar{1}\bar{1}$2] \cite{berbezier_TEM}. 
This gives rise to a pseudohexagonal surface symmetry \cite{chevrier_alpha_rheed,stocker_alpha_Si111}. 
In Fig.\,\ref{fig:surface} the epitaxial relation described by (i) is depicted.
The lattice mismatch (defined as $(a_{Si}-a_{FeSi_2})/a_{Si}$) amounts to 0.79\,\% along Si$[1\bar{1}$0] and 3.92\,\% along Si$[\bar{1}\bar{1}$2].
For simplicity, in the following the directions of the RHEED and NIS measurements, as well as the surface directions of the AFM images are given along the two main Si(111) surface directions, namely Si$\langle\bar{1}$10$\rangle$ and Si$\langle 11\bar{2}\rangle$.

Our RHEED studies confirmed these epitaxial configurations in all samples. Diffraction patterns were recorded along Si$\langle 11\bar{2}\rangle$ [Fig.\,\ref{fig:RHEED}\,(a-g)] and Si$\langle\bar{1}$10$\rangle$  [Fig.\,\ref{fig:RHEED}\,(h-n)]. 
For all substrates a clean 7$\times$7 reconstructed Si(111) surface was confirmed before growth [Fig.\,\ref{fig:RHEED}\,(a,h)]. 
The diffraction pattern along the Si$\langle 11\bar{2}  \rangle$ azimuth [Fig.\,\ref{fig:RHEED}\,(b-g)] consists of the central (222) reflection between the second order (042)/(402) and ($\bar{2}$20)/(2$\bar{2}$0) reflections, accompanied by the intermediate (021) and (201) reflexes [Fig.\,\ref{fig:RHEED}\,(c)] \cite{jedrecy_alpha_TEM}.
For S3, S5 and S6, the main spot of the Si(111) surface is still visible due to the lower $\theta_{Fe}$. 
The RHEED patterns of S1-S4 are dominated by separated diffraction spots suggesting the transmission of the beam through 3D nanoislands. 
On the other hand, the patterns of S5 and especially S6 show a stronger contribution of streaks, which originate from diffraction on crystal truncation rods. 
This indicates the formation of 2D nanoislands with a small extension perpendicular to the Si(111) surface compared to their lateral extension, which is confirmed by the AFM measurements (see below). 
This 3D\,-\,2D transition can also be observed in the RHEED patterns obtained along Si$\langle\bar{1}$10$\rangle$ [Fig.\,\ref{fig:RHEED}\,(i-n)]. 
Along this direction, the diffraction pattern shows two rows of spots consisting of the (001), (002), (003), (004), and (220), (222) reflections \cite{jedrecy_alpha_TEM} [Fig.\,\ref{fig:RHEED}\,(l)], which originate from lattice planes inclined by approximately 35$\,^\circ$ towards the surface. 
When the sample is rotated around the surface normal the inclination angle is repeated every 120\,$^{\circ}$. 
Therefore, it can be concluded that all three possible accommodations of the $\alpha$-FeSi$_2$ unit cell on the Si(111) surface are present in our samples. 
Except for S4, the diffraction patterns show a contribution of the 7$\times$7 reconstructed Si(111)  surface. 
The observed pattern was previously reported for FeSi$_2$ thin films grown on Si(111) by RDE at $T_G$\,=\,500$\,^\circ$C \cite{chevrier_alpha_rheed, minami2002rheed}. 
In \cite{chevrier_alpha_rheed} a grazing-incidence X-ray diffraction study confirms that the investigated structure is surface-stabilized tetragonal $\alpha$-FeSi$_2$ forming the epitaxial relationship to the Si(111) substrate discussed above. 
Furthermore, the observed electron diffraction spots are in agreement with the reciprocal space nodes theoretically predicted for tetragonal $\alpha$-FeSi$_2$  on Si(111) and are in contradiction to the patterns expected for the cubic surface-stabilized \textit{s}- and $\gamma$-phases \cite{jedrecy_alpha_TEM}.

\begin{figure}[t]
\includegraphics[width=0.99\columnwidth]{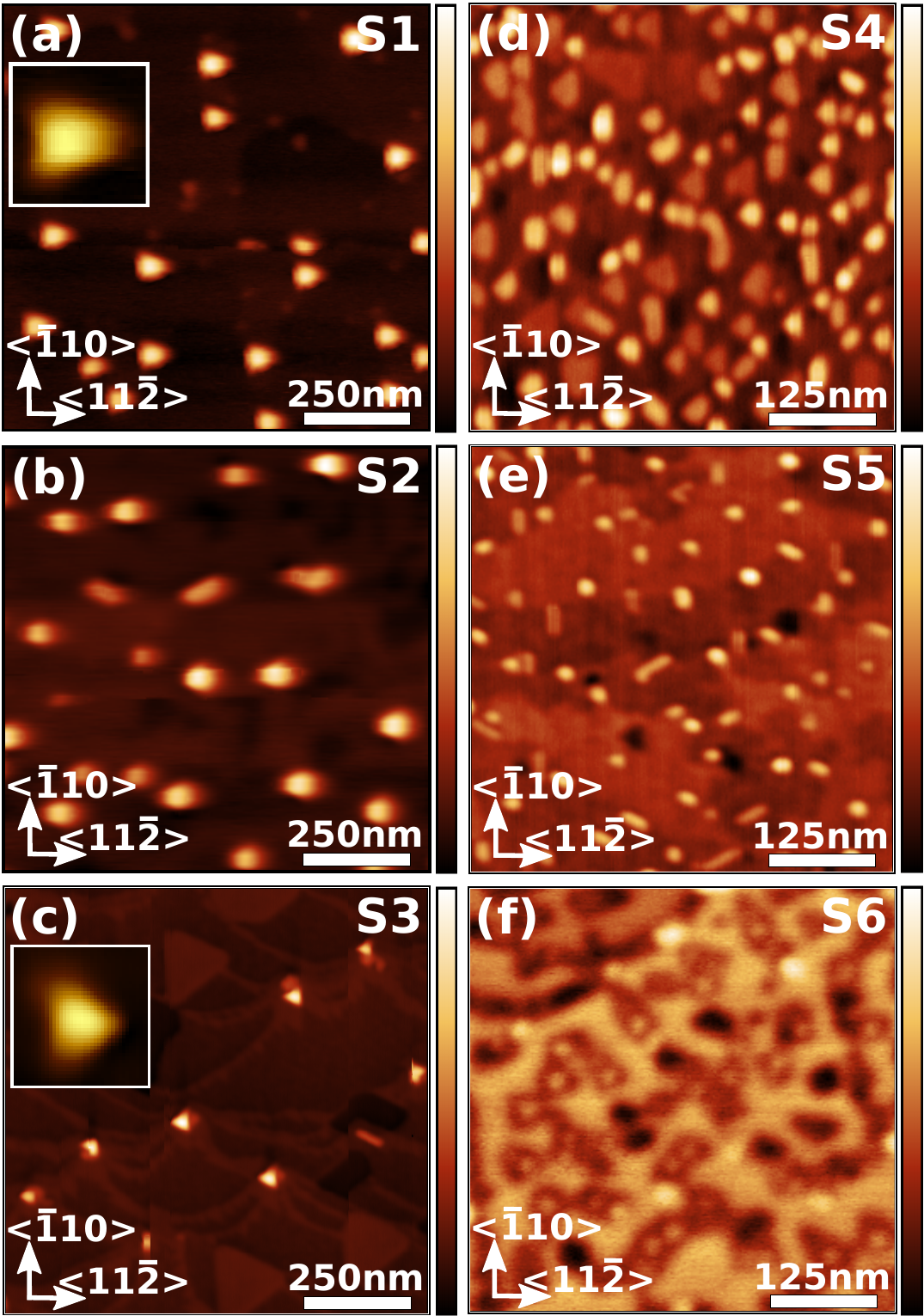}
\caption{
AFM images of samples (a) S1 (height scale 0\,-\,36\,nm), (b) S2 (height scale 0\,-\,34\,nm), (c) S3 (height scale 0\,-\,24\,nm), (d) S4 (height scale 0\,-\,12\,nm), (e) S5 (height scale 0\,-\,11\,nm), and (f) S6 (height scale 0\,-\,9\,nm). The crystallographic directions of the Si(111) surface are indicated. In the insets of (a) and (c) generic islands are depicted.
}
\label{fig:AFM}
\end{figure}

\begin{figure}[t]
\includegraphics[width=0.99\columnwidth]{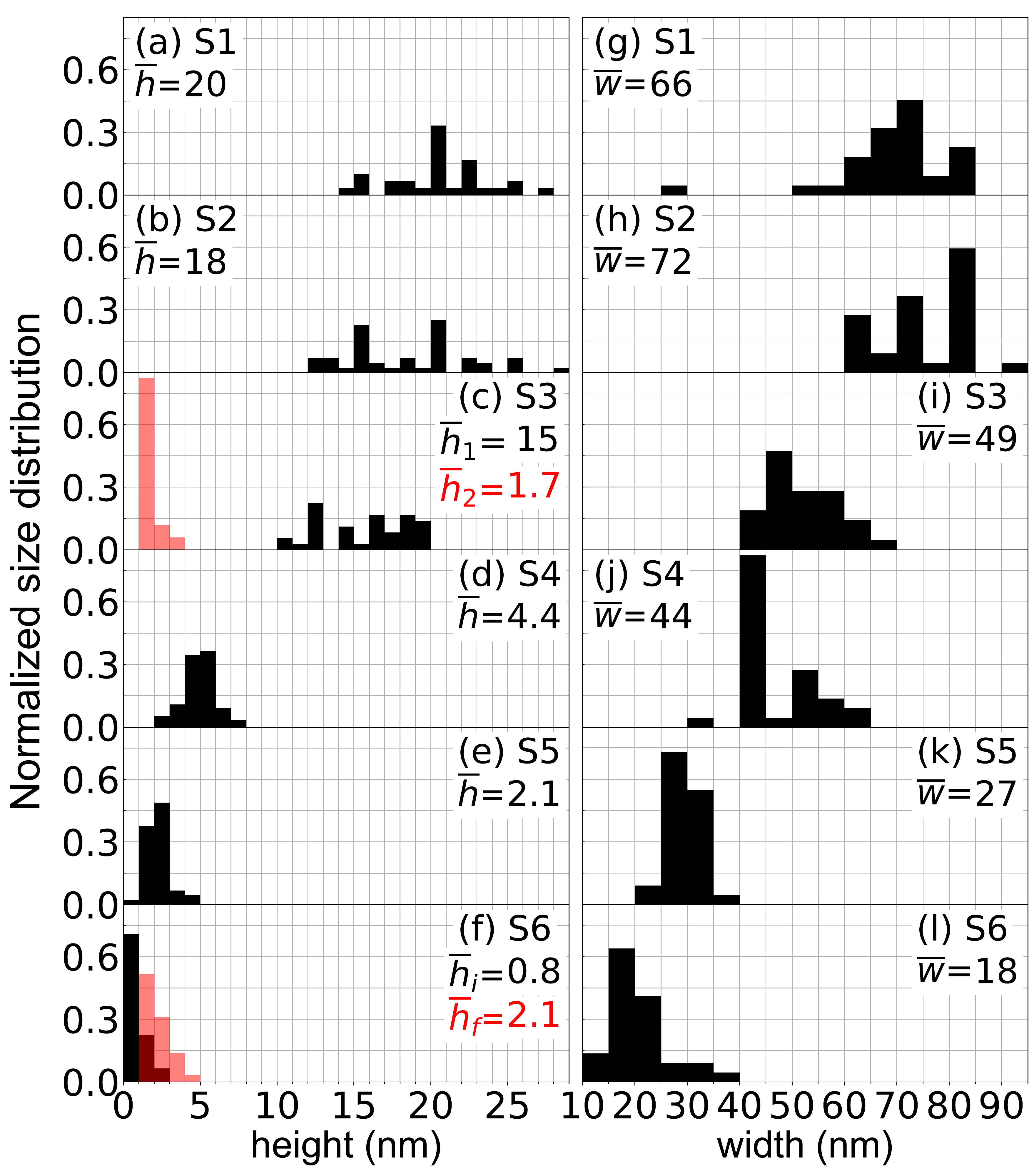}
\caption{Normalized distribution of height (a-f) and width (g-l) of the nanoislands of the indicated samples deduced from the AFM study with the respective average values given in nm. 
For S3 and S6 additionally the height of the flat structures ($\bar{h}_2$) and the film ($\bar{h}_f$), respectively, are given in red. The number of islands measured to obtain the distribution for each sample ranges between 30 and 55.  
}
\label{fig:histogram}
\end{figure}

The surface morphology of the samples was investigated by AFM. 
The results are depicted in Fig.\,\ref{fig:AFM}, in Fig.\,\ref{fig:histogram} the normalized size distribution of the nanoislands obtained from the AFM measurements is shown. 
Fig.\,\ref{fig:AFM} (a) shows the formation of triangular islands for S1 with an average height of $\overline{h}$\,=\,20\,nm and average width of $\overline{w}$\,=\,66\,nm (measured along the symmetry axis of the triangle). 
In case of S2, grown at the same $T_G$ without postgrowth annealing, islands with a slightly decreased average height of $\overline{h}$\,=\,18\,nm and a slightly increased average width of $\overline{w}$\,=\,72\,nm are formed [Fig.\,\ref{fig:AFM}\,(b), Fig.\,\ref{fig:histogram}\,(b,\,h)]. 
A closer look at the AFM image of S2 shows a broadening of the islands along Si$\langle 11\bar{2}  \rangle$, which is the direction of the AFM-tip movement. 
This indicates a tip effect, i.e. a distortion and blurring of the image due to an increased width of the tip used for the measurement of S2. 
This could also be the reason for the spherical shape of the islands, since from the results for S1 and S3 a triangular shape is expected. 
A significant reduction of $\theta_{Fe}$ in S3 compared to S1 and S2 results in a reduction of the average height ($\overline{h}_1$\,=\,15\,nm) and width ($\overline{w}_1$\,=\,49\,nm) [Fig.\,\ref{fig:AFM}\,(c), Fig.\,\ref{fig:histogram}\,(c,\,i)]. 
Similarly to S1, the symmetry axis of the triangular islands is oriented along Si$\langle 11\bar{2}  \rangle$, whereas the edges are pointing along Si$\langle\bar{1}$10$\rangle$. 
A similar orientation of triangular islands on Si(111) has previously been observed for FeSi$_{2}$ \cite{minami2002rheed} and CoSi$_{2}$ \cite{bennett1994CoSi2}. 
In addition, laterally extended ($\overline{w}_2$\,=\,163\,nm) flat ($\overline{h}_2$\,=\,1.7\,nm) structures  are observed. 
When $T_G$ is reduced to 500$\,^\circ$C [S4, Fig.\,\ref{fig:AFM}\,(d) and Fig.\,\ref{fig:histogram}\,(d,\,j)] the height distribution is significantly narrowed with an average value of $\overline{h}$\,=\,4.4\,nm, while the width is only slightly reduced to $\overline{w}$\,=\,44\,nm. 
Despite the higher $T_G$ compared to S4, the lower $\theta_{Fe}$ in case of S5 further narrows the height and the width distribution and the average values are reduced to $\overline{h}$\,=\,2.1\,nm and $\overline{w}$\,=\,27\,nm. 
A combination of low $T_G$ and low $\theta_{Fe}$ [S6,  Fig.\,\ref{fig:AFM} (f) and Fig.\,\ref{fig:histogram} (f,\,l)] leads to the formation of an intermittent FeSi$_{2}$ film along with islands grown in the Si surface areas not covered by the film. 
The average height of the islands is 0.8\,nm, the average height of the film is 2.1\,nm and the average island width is 18\,nm.

In general two growth modes are observed: The samples grown at $T_G\,=\,700\,^\circ$C (S1\,-\,S3) exhibit $\overline{w}$/$\overline{h}$-ratios between 3 and 4, whereas the samples grown at lower temperatures (S4\,-\,S6) form two-dimensional nanoislands with $\overline{w}$/$\overline{h}$-ratios between 10 and 13. 
While the post-growth annealing conducted in case of S1 does not change the morphology significantly compared to S2, a reduction of $\theta_{Fe}$ from 2.2\,\AA  \,(S2) to 0.6\,\AA \,(S3) at $T_G\,=\,700\,^\circ$C leads to the formation of the very flat structures with large lateral extensions, which are only observed at this specific growth conditions. 
Furthermore,  a reduction of $T_G$ by $50\,^\circ$C in case of S3 and S5, both grown with $\theta_{Fe}$\,=\,0.6\,\AA , leads to pronounced changes in the surface morphology and a significantly increased $\overline{w}$/$\overline{h}$ ratio. 
For $\theta_{Fe}$\,=\,2.2\,\AA \, the 3D\,-\,2D transition is observed in the temperature range from $700\,^\circ$C to $500\,^\circ$C. 
These observations indicate  that while $\theta_{Fe}$ clearly influences the morphology, $T_G$ is the more important parameter determining the shape of the nanoislands.

\begin{figure}[t]
\includegraphics[width=0.99\columnwidth]{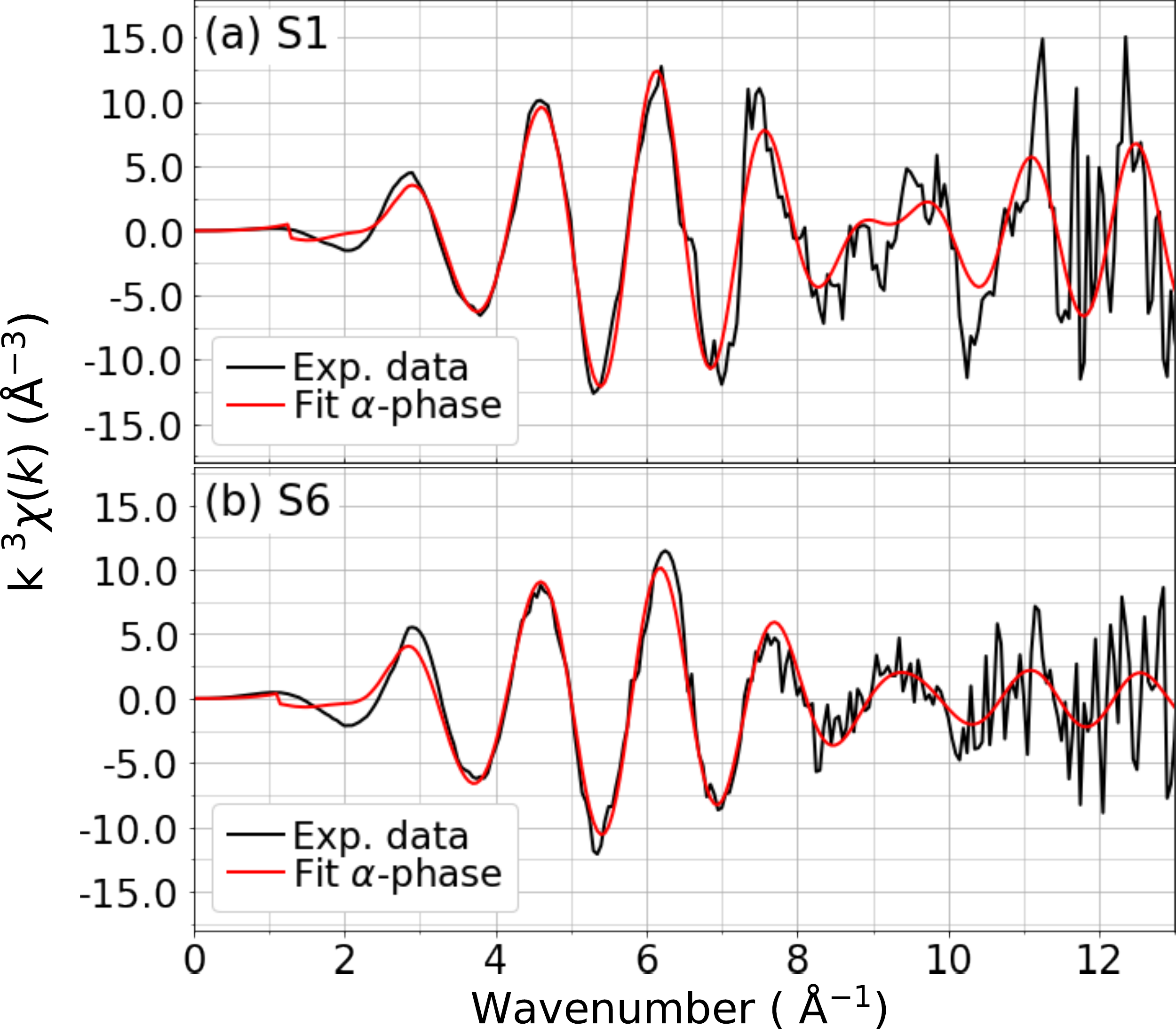}
\caption{Fe\,K-edge EXAFS spectra (black) of (a) S1 and (b) S6  and the respective best fit results (red) obtained by modeling with $\alpha$-FeSi$_2$.
}
\label{fig:EXAFS}
\end{figure}

To validate the information about the crystallographic structure obtained by the RHEED study, the local crystal structure of samples S1-S4 and S6 was investigated by EXAFS measurements. 
In Figure \ref{fig:EXAFS} the representative experimental spectra of S1 and S6 in k space  and the results best fit results are depicted. 
The results of the fits are compared with the expected values for the known FeSi$_2$ phases in Table \ref{tab:tab_EXAFS}. 
Besides $\alpha$-FeSi$_2$ and $\beta$-FeSi$_2$ also the cubic surface-stabilized metallic phases  s-FeSi$_2$ \cite{kaenel_sphase_prb} and $\gamma$-FeSi$_2$ \cite{alvarez_phase_Si111} were considered.
While the interatomic distances obtained from the data analysis are in agreement with the values for the s-phase and $\alpha$-phase, the coordination numbers clearly deviate from the value expected for the s-phase (6 Fe-Fe) and suggest the formation of  $\alpha$-FeSi$_2$ (4 Fe-Fe). 
S6 exhibits a slightly reduced Fe-Fe distance, as well as the lowest Fe-Fe and the highest Fe-Si coordination number. 
FeSi$_2$ nanostructures typically exhibit surfaces with a Si content above the stoichiometric value \cite{sirotti_Si_segragation}.
Therefore, the variations in the coordination numbers can be explained by the very low height of the nanostructures in S6, which leads to the  largest relative amount of atoms located at the FeSi$_2$/Si interface. 
The absorption edges of all Fe K-edge X-ray absorption near edges structure (XANES) spectra, defined as the first inflection point of the rising absorption, are at the energy position for metallic Fe, confirming the metallic nature of the islands. 
An overview of the measured XANES spectra and the Fourier transform of the EXAFS spectra shown in Fig\,\ref{fig:EXAFS} are included in the supplementary material \cite{suppl}.

\begin{table}[t]
\renewcommand{\arraystretch}{1.2}
\begin{tabular}{|C{1.2cm}|C{1.4cm}|C{1.4cm}|C{1.3cm}|C{1.3cm}|C{1.2cm}|}\hline
	&	k-range (\AA$^{-1}$)	&	scattering \linebreak path	& \vspace{0.1mm}  $\sigma^2$  (\AA $^2$) &	coord. \linebreak  number 	& \vspace{0.1mm} d (\AA)		\\
\hline
\multirow{2}{*}{S1}	 & \multirow{2}{*}{3.8\,--\,12.6}					
& Fe-Si 	& 0.004(1) 			 &			6.8(7)					&				2.36(1)					\\
& & Fe-Fe	& 0.0011			 &			3.7(6)					&				2.69(1)					\\
\hline
\multirow{2}{*} {S2}	 & \multirow{2}{*} {3.8\,--\,15.5}					
& Fe-Si     &	0.0047(3) 			 &			8.2(2)					&				2.36(1)					\\
& & Fe-Fe	&	0.0012		 &			3.9(2)					&				2.69(1)					\\
\hline
\multirow{2}{*} {S3}	 & \multirow{2}{*} {3.8\,--\,12.6}					
& Fe-Si 	&	0.007(2)			 &			6.7(7)					&				2.36(2)					\\
& & Fe-Fe	&	0.0012		 &			3.1(11)					&				2.70(2)					\\
\hline
\multirow{2}{*} {S4}	 & \multirow{2}{*} {3.8\,--\,12.6}					
& Fe-Si 	&	0.005(1)		 &			7.7(9)					&				2.36(1)					\\
& & Fe-Fe	&	0.0012			 &			3.7(6)					&				2.69(2)					\\
\hline
\multirow{2}{*} {S6}	 & \multirow{2}{*} {3.8\,--\,12.6}					
& Fe-Si 	&	0.007(2)		 &			8.5(13)					&				2.35(2)					\\
& & Fe-Fe	&	0.0012		    &			2.7(8)					&				2.67(3)					\\
\hline
\multirow{2}{*} {$\alpha$-phase}& \multirow{2}{*} {-}					
& Fe-Si 	&				&			8					&				2.36					\\
& & Fe-Fe	&				&			4					&				2.70					\\
\hline
\multirow{2}{*} {$\beta$-phase}	& \multirow{2}{*} {-}					
& Fe-Si 	&				&				8				&				2.34					\\
& & Fe-Fe	&				&				2 				&				2.97					\\
\hline
\multirow{2}{*} {s-phase}	& \multirow{2}{*} {-}					
& Fe-Si 	&				&			8					&				2.34					\\
& & Fe-Fe	&				&			6					&				2.7					\\
\hline
\multirow{2}{*} {$\gamma$-phase}	& \multirow{2}{*} {-}					
& Fe-Si 	&				&			8					&				2.34					\\
& & Fe-Fe	&				&			12					&				3.84					\\
\hline

\end{tabular}
\caption{Debye-Waller factor ($\sigma^2$), coordination numbers and interatomic distances (d) obtained from modeling of the experimental EXAFS spectra  and theoretical values for the expected FeSi$_2$ phases. 
The $\sigma^2$ values for Si are derived from the EXAFS, whereas the values for Fe are obtained from the NIS experiment. 
The k-range corresponds to the modeled range of the experimental EXAFS data. The values for $\alpha$- and $\beta$-phase are obtained from  ICSD 5257 and 9119, respectively, for  s- and $\gamma$-phase no literature is available.
}  \label{tab:tab_EXAFS}
\end{table}

\subsection{Lattice dynamics} 

\begin{figure}[t!]
\includegraphics[width=0.99\columnwidth]{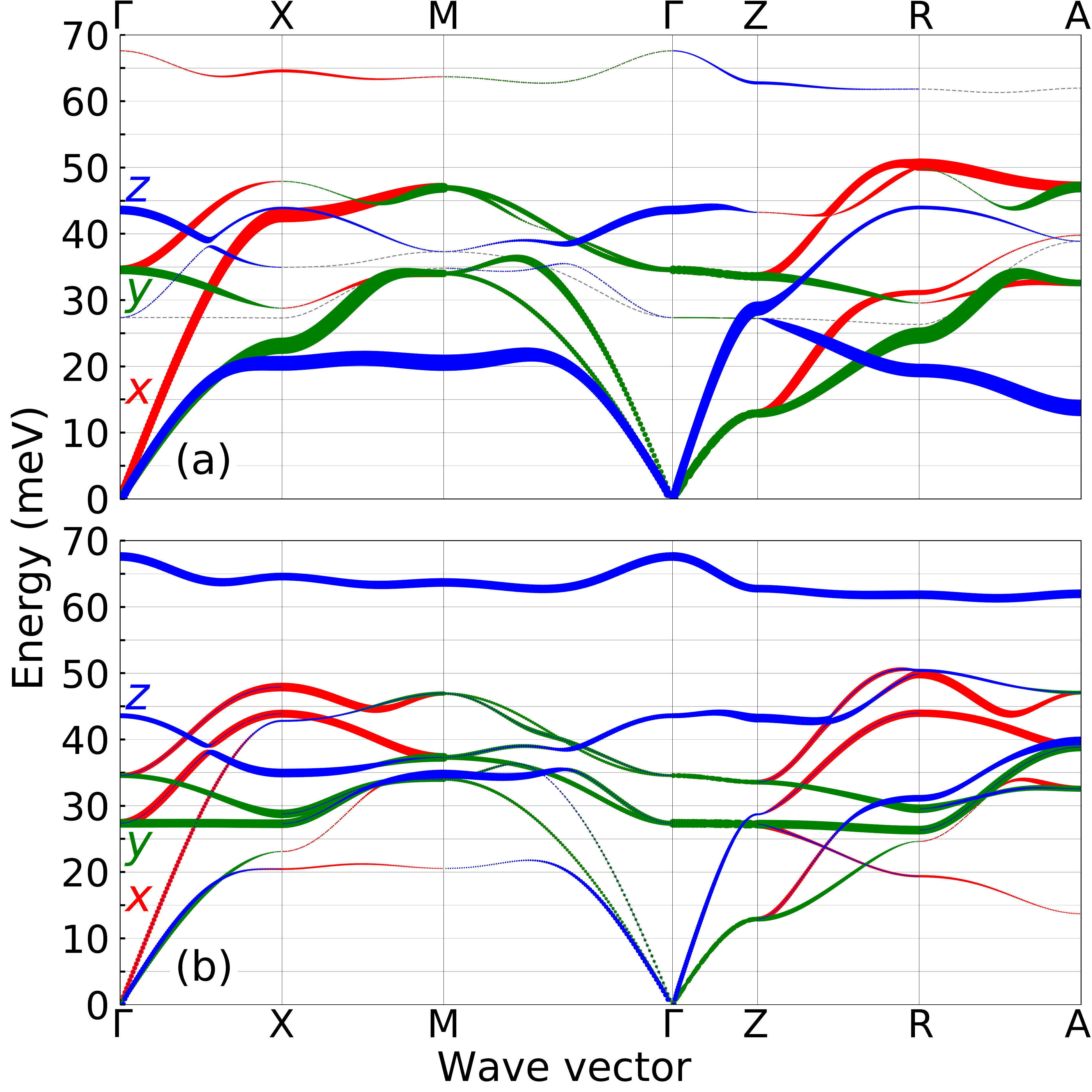}
\caption{\textit{Ab initio} calculated element-specific and polarization-resolved phonon dispersions of (a) the Fe atom and (b) one of the Si atoms in the unit cell of bulk $\alpha$-FeSi$_2$. Red, green and blue lines correspond to  \textit{x}, \textit{y} and  \textit{z} components of the polarization vector, respectively, while the thickness of the lines corresponds to the relative intensity of the given branch. The positions of the high-symmetry points of the Brillouin zone are: {\sffamily \textGamma}(0, 0, 0), {\sffamily X}(0.5, 0, 0), {\sffamily M}(0.5, 0.5, 0), {\sffamily Z}(0, 0, 0.5), {\sffamily R}(0.5, 0, 0.5), {\sffamily A}(0.5, 0.5 0.5). 
}
\label{fig:PDR}
\end{figure}

\begin{figure}[t]
\includegraphics[width=0.99\columnwidth]{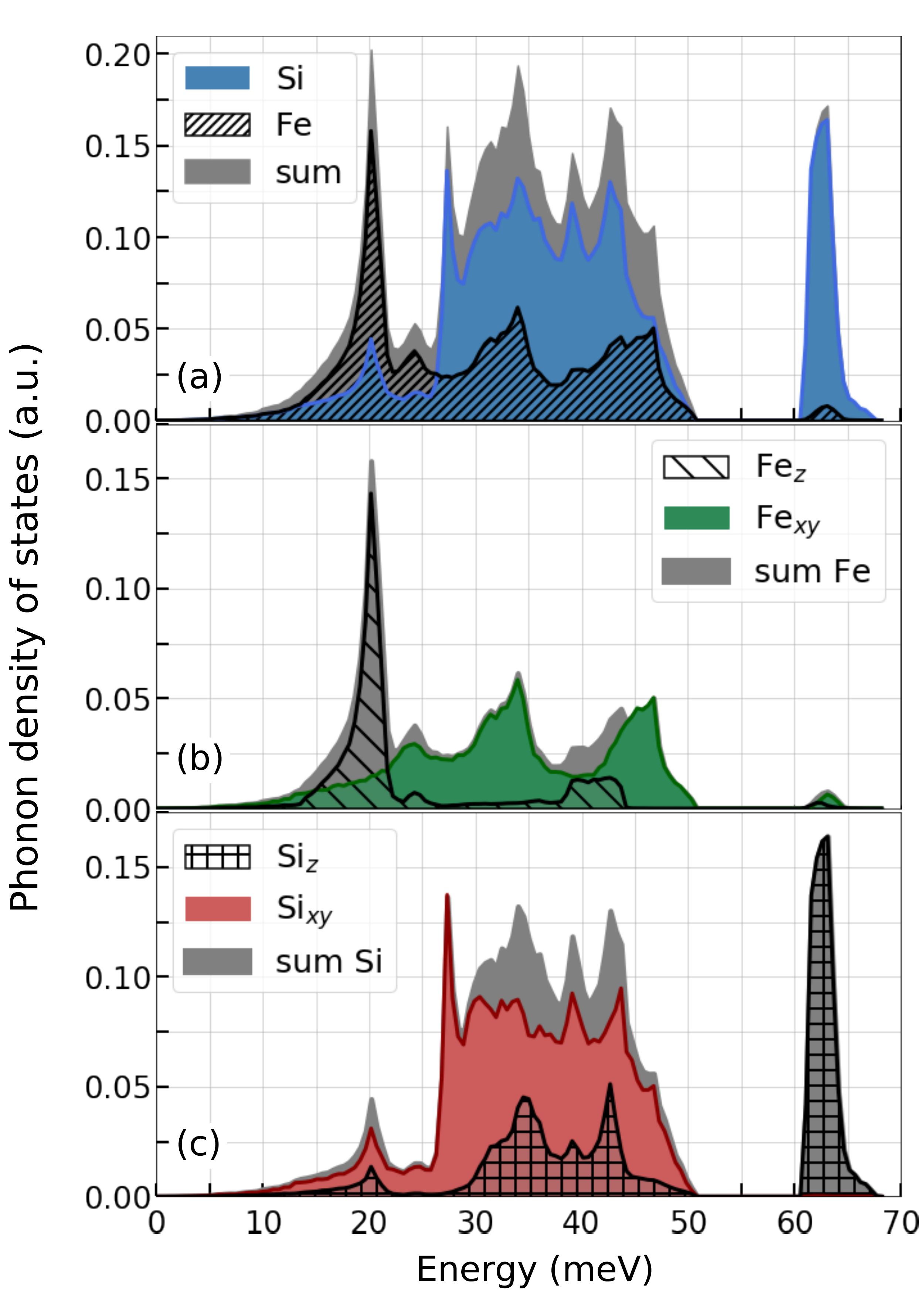}
\caption{(a) \textit{Ab initio} calculated  Fe-partial, Si-partial and total PDOS of bulk $\alpha$-FeSi$_2$. (b) Fe-partial PDOS in  \textit{xy}- and  \textit{z}-polarizations and their sum. (c) Si-partial PDOS in  \textit{xy}- and  \textit{z}-polarizations and their sum.
}
\label{fig:abinitio_PDOS}
\end{figure}

Fig.\,\ref{fig:PDR} depicts the \textit{ab initio} calculated polarization-resolved phonon dispersion relations with contributions from (a) the Fe atom and (b) one of the Si atoms in the unit cell of bulk $\alpha$-FeSi$_2$. 
The Fe-partial dispersion relations show an intense \textit{z}-polarized band at low energies, which is flat between the high-symmetry points {\sffamily X} and {\sffamily M} at 20\,meV. 
The faint mode visible at energies above 60\,meV is coupled to the intense \textit{z}-polarized high-energy mode of the Si atom.  
Vice versa, the low-energy  modes of the Si atom couple to the intense Fe modes. 
In Fig.\,\ref{fig:abinitio_PDOS}\,(a) the total and element-specific PDOS, (b) polarization-projected Fe-partial PDOS, and (c) polarization-projected Si-partial PDOS  of $\alpha$-FeSi$_2$ are shown. Fe$_{xy}$ (Si$_{xy}$) denotes the \textit{xy}-polarized PDOS and Fe$_z$ (Si$_z$) the \textit{z}-polarized PDOS for Fe (Si) atoms. 
The total $\alpha$-FeSi$_2$ PDOS is characterized by pronounced peaks at 20\,meV, mainly originating from vibrations of the Fe atoms, and 63\,meV, mainly originating from vibrations of the Si atoms. 
In the intermediate range between 23 and 50\,meV the silicon contribution is dominant. 
The Fe-partial, polarization-projected PDOS reveals a distinct decoupling of the vibrations with \textit{xy}- and \textit{z}-polarization. 
The \textit{z}-polarized phonon modes observed in the dispersion relations constitute the peak at 20\,meV together with a minor plateau around 40\,meV. 
The \textit{xy}-polarized atomic vibrations exhibit a broader spectrum, which dominates the Fe-partial PDOS at higher energies, i.e. between 25 and 50 meV with peaks at 33 and 45\,meV. 
This is in agreement with previous polarization-resolved \textit{ab initio} and experimental lattice dynamics studies of the tetragonal FePt system, which showed that the \textit{z}-polarized vibrations of the Fe atoms (i.e. along the direction of the unit cell with higher interatomic distances) are characterized by lower energies compared to the \textit{xy}-polarized modes  \cite{tamada_FePt_particles_PDOS,couet_FePt_films_PDOS}.
A minor peak occurs in the PDOS at 24\,meV in both polarizations.
For the Si atoms on the other hand, the restoring force acting along [001] (\textit{z}) is higher compared to [100]/[010] (\textit{x}/\textit{y}) due to the arrangement of the Fe atoms. 
Therefore, in the Si-partial, polarization-projected PDOS [Fig.\,\ref{fig:abinitio_PDOS}\,(c)] the \textit{z}-polarized vibrations constitute the high-energy peak at 63\,meV with a small contribution to the spectrum between 10 and 50\,meV. The PDOS below 50\,meV mainly consists of \textit{xy}-polarized vibrational modes.

\begin{figure}[h!]
\includegraphics[width=0.9\columnwidth]{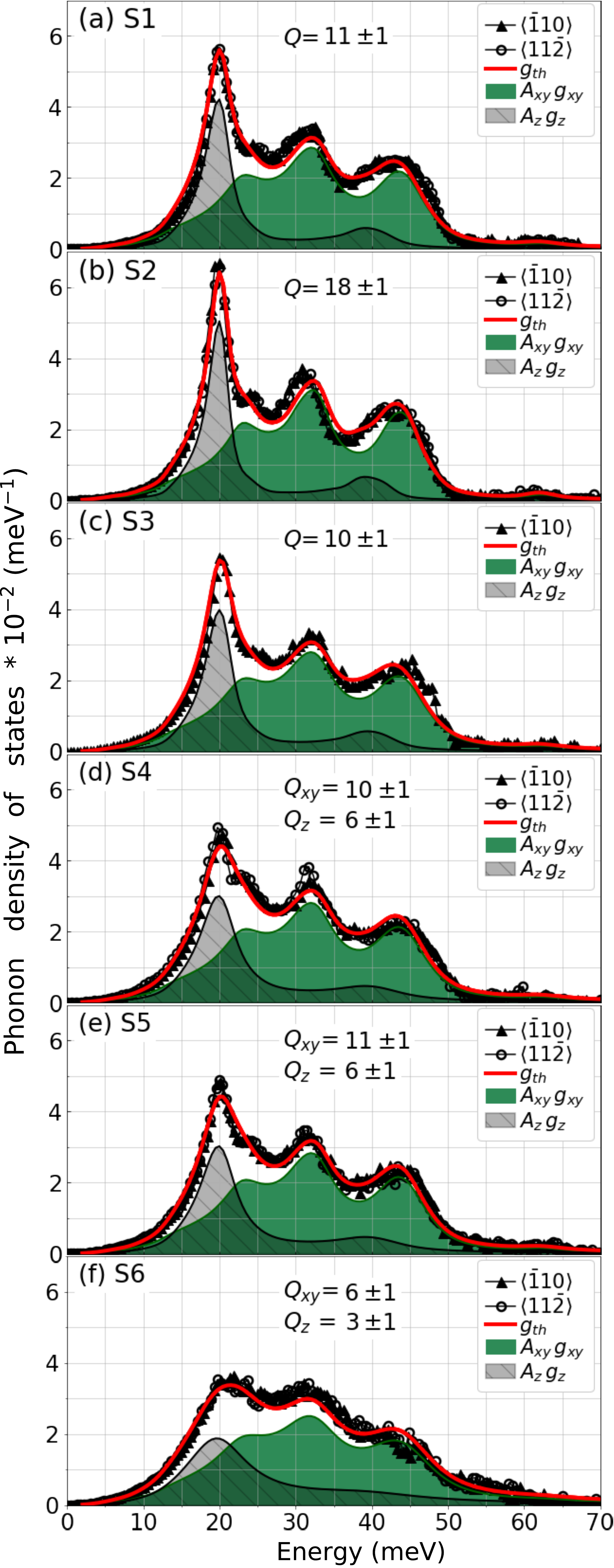}
\caption{ Fe-partial PDOS of the indicated samples measured along Si$\langle\bar{1}$10$\rangle$ and Si$\langle 11\bar{2}\rangle$. The experimental data obtained along Si$\langle\bar{1}$10$\rangle$ is compared with the respective result for $g_{th}$,  which is decomposed into its weighted \textit{xy}\,\,($A_{xy} \, g_{xy}$) and \textit{z}\,\,($A_{z} \, g_{z}$) contributions. The values of $Q$ (S1\,-\,S3), $Q_{xy}$ and $Q_z$ (S4\,-\,S6) obtained from the least squares fit are also given. 
\label{fig:NIS_islands}
}
\end{figure}

Fig.\,\ref{fig:NIS_islands} shows the Fe-partial \cite{comment1} PDOS of S1-S6 measured along the orthogonal directions Si$\langle11\bar{2}\rangle$ and Si$\langle\bar{1}$10$\rangle$ of the Si(111) surface. 
For S3 only the spectrum along Si$\langle\bar{1}$10$\rangle$ was obtained. 
For all investigated samples,  the PDOS measured along the two directions are almost identical and no vibrational anisotropy is observed. 
The peak positions are in good agreement with the \textit{ab initio} calculated Fe-partial PDOS [Fig.\,\ref{fig:abinitio_PDOS}\,(b)].
The main peak of the \textit{z}-polarized vibrations occurs at the predicted position of 20\,meV, whereas for the \textit{xy}-polarized vibrations a small shift of about 1\,-\,2\,meV to lower energies compared to the predicted positions of 33 and 45\,meV is present. 
For S1 and S2, which were measured with higher energy resolution, the minor peak at 24\,meV is visible, whereas in case of S3-S6 a shoulder is observed at a similar position. 
A trace of the peak at 63\,meV is also present in all spectra.
Fig.\,\ref{fig:NIS_islands} also reveals a clear effect of the size of the nanoislands on the shape of the PDOS. 
The peak of the \textit{z}-polarized vibrations at 20\,meV is diminishing with deceasing island size, whereas only in S6 the peaks at 33 and 45\,meV, which originate from the \textit{xy}-polarized vibrations, are significantly affected. 
Furthermore, the number of phonon states below 10\,meV is enhanced by a factor of 1.8 in S6 compared to S1 (for details see \cite{suppl}).

In order to obtain a quantitative understanding of the observed size effect, the experimental PDOS was compared with the \textit{ab initio} calculated polarization-projected Fe-partial PDOS of $\alpha$-FeSi$_2$ considering the crystallographic orientation of the $\alpha$-FeSi$_2$ unit cell on the Si(111) surface. 
According to our RHEED study, three different domain orientations of the $\alpha$-FeSi$_2$ coexist on the Si(111) surface. 
Therefore, the spectrum obtained e.g. with the X-ray wave vector parallel to the Si$\langle\bar{1}$10$\rangle$ azimuth is composed of three spectra measured along different directions of the $\alpha$-FeSi$_2$ crystal, namely [$\bar{1}$10], [$\bar{2}$01] and [0$\bar{2}$1], which are a specific combination of \textit{x}-, \textit{y}- and \textit{z}-polarized phonons \cite{chumakov_anisotropic_nis_FeBO3,kohn_anisotropic_nis_theo}. 
To obtain the relative contributions of the \textit{xy}- and \textit{z}-polarized phonons to the experimental PDOS, the \textit{x}, \textit{y} and \textit{z} vectors of the $\alpha$-FeSi$_2$ unit cell  have to be projected onto the crystallographic directions mentioned above.
This results in relative (\textit{x,y,z}) contributions of (0.3428, 0.3428, 0.3144) along Si$\langle\bar{1}$10$\rangle$ and (0.3432, 0.3432, 0.3136) along Si$\langle11\bar{2}\rangle$, provided that each of the three possible domain orientations has a 1/3 contribution (for details see \cite{suppl}). 
These differences are well below 1\,\% and cannot be resolved in our experiment.

To quantify the strength of the phonon damping, the experimental PDOS obtained along Si$\langle\bar{1}$10$\rangle$ were modeled by  the function $g_{th}(E,Q_{xy},Q_{z})$ defined as:
\begin{equation}
g_{th}(E,Q_{xy},Q_{z})\,=\,A_{xy}\, g_{xy}(E,Q_{xy})+A_{z}\, g_{z}(E,Q_{z})
\label{eq1}
\end{equation} 
with $g_{xy}$ and $g_{z}$ being the \textit{ab initio} calculated \textit{xy}- and \textit{z}-polarized Fe-partial PDOS, respectively, convoluted with the damped harmonic oscillator (DHO) function \cite{Faak} and $A_{xy}$ and  $A_{z}$ being their weighted contributions to the experimental PDOS. 
The DHO function is characterized by a quality factor $Q$  and introduces an energy-dependent broadening of the spectral features  with $Q$ being inversely proportional to the strength of the damping. 
The damping of features in the PDOS is characteristic for nanoscale materials and originates from phonon scattering at atoms located at irregular sites, i.e. defects and dislocations at interfaces and surfaces, as well as within the nanostructure \cite{fultz_dho}.
The DHO function has successfully been used to model and to quantify these effects in nanostructures (see \cite{fultz_dho,Stankov_PRL_Fe,Fe3Si_PRB}).
Taking into account the tensile epitaxial strain induced by the Si substrate,  $g_{xy}$ and $g_{z}$ were calculated assuming a $\alpha$-FeSi$_2$ unit cell with 1\,\% increased lattice parameters in the \textit{xy}-plane (a\,=\,2.72\,\AA, c\,=\,5.14\,\AA).

\begin{figure}[t]
\includegraphics[width=0.99\columnwidth]{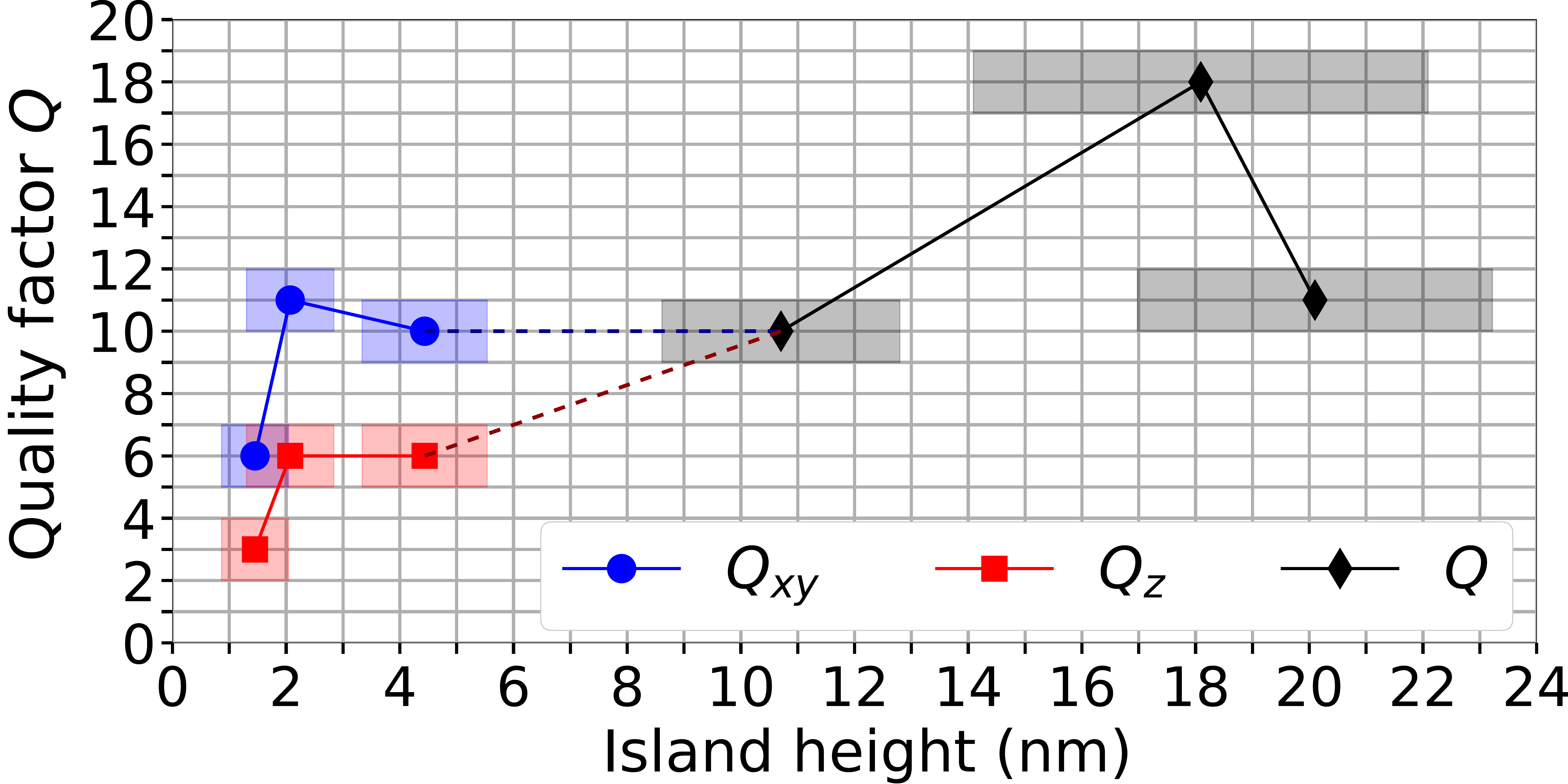}
\caption{Quality factors $Q$ (S1\,-\,S3), $Q_{xy}$ and $Q_z$ (S4\,-\,S6) as a function of the nanoisland height. The error of the height values corresponds to one standard deviation of the height distribution displayed in Fig.\,\ref{fig:histogram}.
}
\label{fig:Q_thickness}
\end{figure}

Prior to modeling, $g_{xy}$ and $g_{z}$ were determined considering the experimental instrumental function of the respective beamline to ensure a valid comparison between the spectra obtained with different energy resolutions \cite{note2}. 
Subsequently, the respective experimental PDOS measured along Si$\langle\bar{1}$10$\rangle$ was fitted with $g_{th}$ (Eq. (1)) using the least-squares method. 
The contributions of the \textit{xy}- and \textit{z}-polarized phonons were fixed to $A_{xy}\,=\,0.69$, $A_{z}\,=\,0.31$, following the discussion of the (\textit{x},\textit{y},\textit{z}) components in the previous paragraph. 
The data was fitted with two different approaches. 
In the first approach, we assumed $Q_{xy}$\,=\,$Q_{z}$, i.e. the same damping for both contributions. 
In the second approach, $Q_{xy}$ and $Q_{z}$ were independent parameters. 
For S1, S2 and S3 both approaches led to very similar results, whereas for S4, S5 and S6 the approach considering a polarization-dependent damping significantly improved the agreement between experiment and theory. 
Fig.\,\ref{fig:NIS_islands} shows the results using one common $Q$ for S1, S2 and S3, whereas for S4, S5 and S6 the results of the fits with $Q_{xy}$ and $Q_{z}$ being independent parameters are displayed.  
In Fig.\,\ref{fig:Q_thickness} the $Q$ values for the respective samples are plotted as a function of the average island height. 
For S6 the average of the height  of the  islands and the film and for S3 the weighted average of 3D islands and flat structures  is used. 
The width of the islands is not considered, since the width/height ratio is between 3 and 4 for S1\,-\,S3 and between 10 and 13 for S4\,-\,S6. 
For this reason, confinement effects are expected to arise primarily due to the reduction of the height of the nanoislands.

The sharper features of the PDOS of S2 result in a significantly higher Q value compared to S1, despite the fact that both samples exhibit similar average sizes and size distributions of the islands (Fig.\,\ref{fig:histogram}).
The low surface-to-volume ratio in the large islands of S1 and S2 implies that ca. 90\,\% of the Fe atoms exhibit a bulk-like coordination. 
Thus, an impact of the capping layer present in S1 or the free surface present in S2 is not expected. 
The observed differences are possibly a result of the post-growth annealing of S1. 
It could reduce the substrate/FeSi$_2$ interface sharpness and therefore increase the number of atoms located at irregular sites. 
For S3, which is grown at the same T$_G$ with lower $\theta_{Fe}$, a clear reduction of the $Q$ value compared to S2 is observed. 
The stronger damping is expected to arise from atoms in the flat structures which coexist with the 3D islands. 
When the height distribution is narrowed and the average height is reduced below 5\,nm in S4, a polarization dependence of the phonon damping is observed. 
While $Q_{xy}$ coincides with the $Q$ value obtained for S3, $Q_{z}$ is significantly reduced. 
Despite the reduction of the average island height and width in S5 compared to S4, the $Q_{xy}$ and $Q_{z}$ values are very similar in both samples. 
This could be a consequence from the higher $T_G$ in case of S5, which leads to a higher degree of crystalline order and therefore to a reduction of the concentration of defects inside the nanoislands, which compensates the size effect \cite{cuenya_Fe_particles}. 
Another reason for the similar quality  factors could be the fact that S4 is measured \textit{in situ} whereas  S5 was capped with Si. 
Due to the high surface-to-volume ratio of the nanoislands in S4 and S5 compared to S1 and S2, the influence of the capping layer on the PDOS of S5 is no longer negligible.
The capping layer could partially suppress the soft phonon modes originating from the broken translational symmetry of the surface and therefore compensate the phonon-damping effect induced by the reduction of structure height. 
For S6, the sample grown at the same $T_G$ as S4 but with lower Fe coverage, resulting in the smallest islands, $Q_{z}$ is again reduced compared to S4 and S5. Also $Q_{xy}$ is significantly reduced compared to the quality factors of all other samples.

\begin{table*}[t]
\begin{center}

\renewcommand{\arraystretch}{1.2}
 \begin{tabular}{|c|c|c|c|c|c|c|c|}
\hline
		     							      & &  $F\,(N/m)$	& $\langle x^2 \rangle$\,(\AA$^2$)	& $S_V$\,($k_B/atom$) & $C_V$\,($k_B/atom$) &$\alpha$	($10^{-5}meV^{-3}$)	& $v_S$    $(m/s)$	\\ 
\hline 																																			   	
\hline
\multirow{4}{1em} {\rotatebox{90}{theory}}	 
&\textit{xy}	          		                &  271 		 	& 0.0094							& 2.52 					& 2.57				& &				5220								\\
&\textit{z}	          		                	&  146		 	& 0.0141							& 3.45					& 2.76				& &				5430							\\
&sum  $\langle 11\bar{2}\rangle$ proj.   		&  232		 	& 0.0109							& 2.81 					& 2.63				& &											\\
&sum  $\langle\bar{1}$10$\rangle$ proj.   		&  232		 	& 0.0109 							& 2.81 					& 2.63				& &											\\
\hline 																																			   	
\hline
\multirow{11}{1em} {\rotatebox{90}{experiment}}	 
&S1 	 $\langle 11\bar{2}\rangle$			 		& 230(5)	 	& 0.0110(2)							& 2.84(2)				& 2.63(2)			&	2.81(1)	&  4903(10) 						\\
&S1 	 $\langle\bar{1}$10$\rangle$ 				& 233(5)	 	& 0.0108(2)							& 2.83(2)				& 2.63(2)			&	2.77(1)	&  4923(10)					\\
\cline{2-8}
&S2 	 $\langle 11\bar{2}\rangle$					& 223(5)	 	& 0.0114(2)							& 2.92(2)				& 2.65(2)			&   2.89(1)	&  4855(12)					\\
&S2 	 $\langle\bar{1}$10$\rangle$				& 217(5)	 	& 0.0117(2)							& 2.95(2)				& 2.66(2)			&	2.90(2)	&  4848(9)					\\
\cline{2-8}
&S3 	$\langle\bar{1}$10$\rangle$					& 233(5)	 	& 0.0115(2)							& 2.85(2)				& 2.63(2)			&   3.93(2)	&  4383(8)					\\
\cline{2-8}
&S4 	 $\langle 11\bar{2}\rangle$ 				& 214(5)	 	& 0.0123(2)							& 2.96(2) 				& 2.66(2) 			&	4.02(2)	&  4349(7)						\\
&S4 	 $\langle\bar{1}$10$\rangle$ 				& 227(5)	 	& 0.0119(2)							& 2.90(2)				& 2.64(2) 			&	3.87(2)	&  4404(7)						\\
\cline{2-8}
&S5 	 $\langle 11\bar{2}\rangle$ 				& 234(5)	 	& 0.0119(2)							& 2.87(2) 				& 2.63(2) 			&	4.27(3)	&  4262(10)					 \\
&S5 	 $\langle\bar{1}$10$\rangle$ 				& 232(5)	 	& 0.0116(2)							& 2.85(2)				& 2.63(2) 			&	3.90(4) &  4394(11)					\\
\cline{2-8}
&S6 	 $\langle 11\bar{2}\rangle$					& 238(5)	 	& 0.0122(2)							& 2.84(2) 				& 2.62(2) 			&	4.90(2)	&  4072(6)					 \\
&S6 	 $\langle\bar{1}$10$\rangle$ 				& 236(5)	 	& 0.0122(2)							& 2.85(2)				& 2.63(2) 			&	4.94(1)	&  4061(5)					  \\
\hline
 \end{tabular}
 \caption{Fe-partial mean force constant $F$, mean square displacement $\langle x^2 \rangle$, vibrational entropy $S_V$ and heat capacity $C_V$ calculated from the \textit{ab initio} \textit{xy}- and \textit{z}-polarized PDOS for  $\alpha$-phase FeSi$_2$, from their weighted sum projected along $\langle 11\bar{2}\rangle$ and $\langle\bar{1}$10$\rangle$, as well as from the experimental PDOS. The coefficient $ \alpha$ derived from the low-energy part of the PDOS ($g(E)\,=\,\alpha E^2$)  and the sound velocity $v_S$ are also given.
}  \label{tab:TDP_sound}
\end{center}
\end{table*}

Metal-silicide heterostructures grown by reactive deposition or solid phase epitaxy exhibit an intrinsic degree of disorder at the Si/silicide interface \cite{berbezier_TEM}.
Moreover, the atoms at the surface of the islands experience a broken periodicity due to the low coordination in case of the \textit{in situ} measured samples or the transition to the amorphous Si layer in case of the capped samples. 
Therefore, at both interfaces the lifetime of the phonons is reduced and the width of the respective PDOS features is increased. 
For nanoislands grown at the same temperature with different $\theta_{Fe}$ it can be assumed that the surface-to-volume ratio is higher for smaller $\theta_{Fe}$ (e.g. in S4 and S6).
Consequently, the relative fraction of atoms at the surface/interface is increased, leading to significantly smaller quality factors.

Furthermore, we observe an onset of polarization-dependent damping of the PDOS features in the height range from 10\,-\,5\,nm. 
This can also directly be seen in the experimental spectra as the shape of the PDOS features of the \textit{xy}-polarized phonons at 33 and 45\,meV does not significantly change from S1 to S5 whereas the peak at 20\,meV exhibits a clear dependence on the nanoislands height. 
One possible explanation for the stronger damping of the \textit{z}-polarized phonons is that they mostly consist of transverse,  low-energy vibrations. 
Therefore, a coupling to the soft modes present at surfaces \cite{slezak_PRL} and interfaces \cite{Fe3Si_PRB} is more likely than it is for the \textit{xy}-polarized vibrations, which exhibit higher energies. 
Thus, phonons polarized along \textit{z}-direction are more sensitive to the effects induced by nanoscaling of the  $\alpha$-FeSi$_2$ crystal.

To examine the validity of our results, which are based on the assumption that the $A_{xy}$ and $A_z$ values can be determined by the epitaxial relations, the experimental data was additionally modeled with $A_{xy}$ and $A_{z}$ being free parameters in the mean square optimization. 
In this case the values of $Q$ for S1, S2 and S3 as well as the $Q_{xy}$ and $Q_{z}$ values for S4, S5 and S6 coincided within the uncertainty  and the $A_{xy}$ values are on average only slightly increased by 4\,\% compared to the theoretically predicted value of $A_{xy}$=0.69.

\subsection{Thermodynamic and elastic properties} 

The thermodynamic and elastic properties derived from the Fe-partial \textit{xy}- and \textit{z}-polarized \textit{ab initio} calculated PDOS [Fig.\,\ref{fig:abinitio_PDOS}\,(b)] and their weighted sum projected along Si$\langle 11\bar{2}\rangle$ and Si$\langle\bar{1}$10$\rangle$, as well as the values calculated from the experimental PDOS of S1-S6 (Fig.\,\ref{fig:NIS_islands}) are presented in Table \ref{tab:TDP_sound}. 
The coefficient $ \alpha$ of the Debye model ($g(E)\,=\,\alpha E^2$) and the sound velocity $v_S$ are also given. 
In accordance with the vibrational anisotropy evidenced in Fig.\,\ref{fig:abinitio_PDOS}(b), the \textit{xy}- and \textit{z}-projected values of the thermoelastic properties calculated from Foretical PDOS differ significantly. 
As a consequence of the higher interatomic distances of the Fe atoms along \textit{z}-direction in the tetragonal  $\alpha$-FeSi$_2$ unit cell,  the force constant $F$ is reduced by 46\,\% compared to the \textit{xy}-direction.  
This results in a mean square displacement along \textit{z}-direction increased by 50\,\% compared to the value obtained along \textit{xy}-direction. 
Furthermore, the vibrational entropy $S_V$ is higher by 37\,\% and the heat capacity $C_V$ by 7\,\%  along \textit{z}-direction, compared to \textit{xy}-direction. 
A projection of the PDOS along Si$\langle 11\bar{2}\rangle$ and Si$\langle\bar{1}$10$\rangle$ leads to a slightly lower contribution of the \textit{z}-polarized vibrational modes since the relative \textit{z}-contribution is reduced from $A_z^{bulk}$\,=\,0.33 to $A_z$\,=\,0.31.
A comparison of the experimental and the \textit{ab initio} calculated values projected along the respective directions does not show a systematic development of the mean force constant. 
With a decrease of 8\,\%  S4\,$\langle\bar{1}$10$\rangle$ exhibits the biggest deviation. 
While S1 shows very good agreement with the theoretical values for the mean square displacement, the reduction of island height leads to an enhancement of 12\,\% in S6 compared to S1.
For $S_V$ the experimental values are on average increased by 2.4\,\%, while for  $C_V$ the average increase is below 0.3\,\%.

The low-energy part of the PDOS can be described by the Debye model ($g(E)\,=\,\alpha E^2$). 
The coefficient $\alpha$ is derived from the low-energy region of the experimental data and shows a clear trend towards higher values for decreasing island size from S1 to S6. 
The value for S6\,$\langle\bar{1}$10$\rangle$ is enhanced by a factor of 1.8 compared to S1\,$\langle\bar{1}$10$\rangle$. 
In the 2D structures the number of low-energy states is increased due to lower coordination \cite{slezak_PRL}, interface-specific phonon states \cite{Fe3Si_PRB}, or epitaxial strain induced by lattice mismatch \cite{Stankov_PRL_Fe}. 
A deviation of $g(E)$ from the quadratic energy dependence is not observed in our experiment, as it is reported for iron nanoclusters with diameters of about 10\,nm \cite{cuenya_Fe_clusters}. 
In \cite{cuenya_Fe_clusters} it is attributed to the low coordination of atoms located at the surface. 
Very likely the reason for this different behavior is the fact that the nanostructures grown by RDE are strongly coupled to the substrate.

The theoretical values for the sound velocity $v_S$ were calculated  from the slope of the acoustic branches of the phonon dispersions, while the experimental values were determined  using the coefficient $\alpha$  \cite{hu_sos}. 
In the islands of S1 and S2 the $v_S$ values are on average 8\,\% below the theoretically predicted numbers. 
The differences can be explained by the fact that the calculations are performed for a perfect crystal, whereas in the nanoislands of S1 and S2 the propagation of sound waves is decelerated by scattering on interfaces. 
Due to the higher surface-to-volume ratio in the smaller nanoislands, $v_S$ is reduced by 18\,\% in S6 compared to S1.
The theoretical results for $v_S$ obtained here are smaller from those reported in \cite{wu_FeSi_prop}.

\section{Conclusions}\label{Conclusions}
FeSi$_2$ nanoislands of the surface-stabilized $\alpha$-phase were grown on Si(111) via reactive deposition epitaxy.
The previously reported epitaxial relationship between the substrate and the $\alpha$-FeSi$_2$ was confirmed by RHEED. 
An EXAFS study proved the formation of $\alpha$-FeSi$_2$ and excluded other known surface-stabilized phases.
The surface morphology was investigated via AFM.
The average height of the islands was in the range from 1.5 to 20\,nm and the average width from 18 to 72\,nm. 
Two growth regimes were observed: at $T_G\,=\,700\,^\circ$C mostly 3D nanostructures with width/height ratios between 3 and 4 are formed, whereas lower growth temperatures led to the formation of 2D nanostructures with width/height-ratios between 10 and 13.

The lattice dynamics  of $\alpha$-FeSi$_2$ was determined experimentally with nuclear inelastic scattering performed at room temperature and first-principles calculations of the polarization-projected, element-specific phonon dispersions and phonon density of states.
The measurement of the Fe-partial PDOS along two orthogonal directions on the Si(111) surface, namely Si$\langle\bar{1}$10$\rangle$ and Si$\langle 11\bar{2}\rangle$, revealed:
(i) a vibrational isotropy, despite of the strong anisotropy of the unit cell of the tetragonal $\alpha$-FeSi$_2$; 
(ii) a pronounced size- and phonon polarization-dependent behavior. 
The first observation is explained by the three different domain orientations of the $\alpha$-FeSi$_2$ on the Si(111) surface.
The PDOS resulting from these domain orientations  projected  along Si$\langle\bar{1}$10$\rangle$ and Si$\langle 11\bar{2}\rangle$ are almost identical and therefore vibrational isotropy is observed.
Modeling of the experimental data with the \textit{ab initio} calculations shows that the reduction of the height of the nanoislands results in a damping of all phonon peaks, being particularly strong for the low-energy \textit{z}-polarized phonons for average islands heights below 10\,nm.
This effect is explained by the lower energy of the \textit{z}-polarized phonons, compared to the \textit{xy}-polarized vibrations, which results in a more efficient coupling to the low-energy surface/interface vibrational modes.

The vibrational anisotropy of $\alpha$-FeSi$_2$ revealed by the \textit{ab initio} calculations is reflected in the thermodynamic properties.
The theoretical value of the mean force constant along \textit{z}-direction is reduced by 46\,\% compared to the \textit{xy}-plane, while the mean square displacement, vibrational entropy and lattice heat capacity  are increased by 50\,\%, 37\,\% and 7\,\%, respectively.
The reduction of the height of the nanoislands leads to an increase of the mean square displacements by 12\,\% and a decrease of the sound velocity by 18\,\% in the smallest islands.

The reported results demonstrate that atomic vibrations along the crystallographic directions characterized with lower mean force constant, which exhibit in general lower energies, couple more efficiently to low-energy surface/interface vibrational modes. 
The observed phonon polarization-dependent damping in nanostructures should be generally valid for single-crystalline materials with non-cubic unit cells. \\

\begin{acknowledgments}
S.S. acknowledges the financial support by  the  Helmholtz  Association  (VH-NG-625)  and  BMBF (05K16VK4).  P.P. acknowledges support by the Narodowe Centrum Nauki (NCN, National Science Centre)
under Project No. 2017/25/B/ST3/02586 and the access to ESRF financed by the Polish Ministry of Science and High Education, decision number: DIR/WK/2016/19. The European Synchrotron Radiation Facility is acknowledged for beamtime provision at the Nuclear Resonance beamline ID18. We thank Mr. J.-P. Celse for technical assistance during the experiment at ID18. We acknowledge DESY (Hamburg, Germany), a member of the Helmholtz Association HGF, for the provision of experimental facilities. Parts of this research were carried out at the High Resolution Dynamics Beamline P01 at PETRA III. We thank Mr. C. Hagemeister and Mr. F.-U. Dill  for technical assistance during the experiment at P01.
\end{acknowledgments}


\begin{thebibliography}{1}
\bibitem{comment1}   Nuclear inelastic scattering is sensitive only to the atomic vibrations of the M\"ossbauer-active isotope, accordingly the method gives access to the partial PDOS of the $^{57}$Fe atoms in the $\alpha$-FeSi$_2$  nanostructures.
\end{thebibliography}
\end{document}